\documentstyle[12pt]{article}

\newcommand{\G}{{\cal G\mit}}

\newtheorem{th}{Theorem}[section]
\newtheorem{defin}[th]{Definition}

\newtheorem{lemma}[th]{Lemma}

\begin{document}

   \begin{titlepage}
          \title{
    {\bf  Racks and orbits of dressing
                transformations  }       }

      \author{ \\ {\bf
       A. A. Balinsky} \thanks{e-mail:
       balin@leeor.technion.ac.il }  \\ \\
   Technion-Israel Institute of Technology \\
      Department of Mathematics \\
       32000 Haifa, ISRAEL}

   \date{ }

   \maketitle

 \thispagestyle{empty}

  \begin{abstract}
         New algebraic structure on
         the orbits of dressing
         transformations of the
         quasitriangular Poisson Lie
         groups is provided.
         This give the topological
         interpretation
         of the link invariants
         associated with the
         Weinstein--Xu classical solutions
         of the quantum Yang-Baxter
         equation. Some applications
         to the three-dimensional
         topological quantum
         field theories are discussed.

          \end{abstract}

      \end{titlepage}


\section*{Introduction.}
Three-dimensional topological quantum
field theories and especially
Chern--Simons type theory
(see \cite{At,Wit,AC,MR})
have been attracting interest of
mathematicians and physicists.
Many of them give us the new invariants
of links and 3-manifolds.
  The article
  \cite{JB} gives an excellent account
  of main developments in
  knot theory which followed upon
  the discovery of the Jones
  polynomials \cite{VJ} in 1984.
  But from our point of view, we
  nevertheless are far from the
  understanding of the topological
  meaning of the new invariants.

In the very deep paper \cite{WX}
A. Weinstein and P. Xu defined a
broad class of knot and links
invariants using a kind of the
classical solutions of the
quantum Yang--Baxter equation. In
\cite{WX} the classical analogue
is developed for part of the
standard construction in which
generalized Jones invariants
are produced from representations
of quantum groups. As far as
I know their article is the
first attempt to  understand
the topological meaning of
the {\em quantum group
invariants} on the quasi-classical level.
It was established in \cite{BAL1} that
in the case of
factorizable Poisson  Lie group
$G$ the Weinstein-Xu link invariant
coincide with the
space of  link group representations
in $G$. The general case of an arbitrary
quasitriangular Poisson Lie group
is connected with Joyce's
theory of knot quandles
or fundamental racks.

Any codimension two link has a
fundamental rack which contains
more information than the fundamental
group and which is a complete
invariant for irreducible links
in any 3-manifold.

Rack provides complete algebraic and
topological framework in which
to study links and knots in 3-manifolds.
The finding rack structures inside
of the quantum link invariants
and of the three-dimensional
topological quantum
field theories looks like the
finding hidden symmetries
in the integrable equations.
It was done in \cite{Yett}
  for the topological quantum
field theories associated to
finite groups.
I think that this is a very
perspective area for the investigation
and that the concept of rack gives us the
powerful tools for description of the
topological quantum
field theories.

Our main goal in this paper is to
give the rack structure for
Poisson Lie group. This is the main
ingredient in our interpretation
of the Weinstein-Xu link invariant
in the general case of an arbitrary
quasitriangular Poisson Lie group.

 \section{Racks and Quandles}
In this section we state
 some properties of the racks and quandles
which will be used in this paper.
For more details on this subject,
see \cite{FR,Joyce,BR}.
To simplify reading, we keep the
notations of \cite{FR}  wherever possible,
on one hand, but give all
necessary definitions, on the other.

Recall that a set with product is a
pair $(\Delta, \ast)$ where
$\Delta$ is a set and $\ast$ is a
map $\Delta \times \Delta
\rightarrow \Delta$. The value of this
map for $(a,b)$ will be
denoted by $a^{b}$ or by $a \ast b$.
The reasons for writing the
operation exponentially
are explained in \cite{FR}.
A morphism of sets with product $(\Delta,
\ast) \rightarrow (\Delta ', \ast ')$ is a
map $\phi : \Delta
\rightarrow \Delta '$ such that
$\phi (a \ast b) = \phi (a) \ast
' \phi (b)$. For any set with product
$(\Delta, \ast)$ and $b \in
\Delta$ the right translation $r_{b}$
is the map
$r_{b} : \Delta \rightarrow \Delta $
defined by
$r_{b} (a) = a \ast b = a^{b}$.

\begin{defin}[\cite{FR}] A {\bf rack}
is a non-empty set $\Delta$ with a product
satisfying the following two axioms:
\begin{itemize}
\item Given $a,b \in \Delta$
there is a unique $c \in \Delta$ such
that $a=c^{b}.$
\item  Given $a,b,c \in \Delta$
the formula
\[a^{bc}=a^{cb^{c}} \]
holds.
\end{itemize}
Here  $a^{bc}$ means $(a^{b})^{c}$ and
$a^{b^{c}} $ means $a^{(b^{c})}$.
\end{defin}

In other words, a  set  with
product $(\Delta, \ast)$
is a rack, iff
all right translations are automorphisms:
\begin{itemize}
\item  $\forall a,b \in
\Delta \ \ \ \ \exists ! \ \ \ \ c \in
            \Delta \ \ \ \ a=c \ast b $
\item $\forall a,b,c \in \Delta \ \ \ \ \ \ \ \
                 (a \ast c) \ast (b \ast c)
 = (a \ast b) \ast c $.
\end{itemize}
One can find many examples
of  automorphic sets in \cite{FR,BR}.

The rack axioms are the algebraic distillation
of two of the Reidemeister moves
(the second and third moves).

\begin{defin}[\cite{WX,DP}]

A map $R : S \times S \rightarrow S \times S$,
 where $S$ is any set, is called a
 solution to the set-theoretic quantum
Yang--Baxter equation if
\[   R_{13}  R_{23} R_{12}  = R_{12} R_{23} R_{13}, \]
where $R_{ij} : S \times S \times S \rightarrow S \times S \times S$
is $R$ on the $i^{th}$  and $j^{th}$  factors of the
cartesian product and $Id$ on the third one.

\end{defin}
The following fact is crucial
in the using of racks in low-dimensional
topology.
\begin{lemma}
If  $\Delta$ is an   rack
 then the map
\[ R : \Delta \times \Delta
\rightarrow \Delta \times \Delta
\ \ \ \ \ \  (a,b) \mapsto (a, b^{a}) \]
 is a solution to the
set-theoretic Yang-Baxter equation.
\end{lemma}

Given a rack  $\Delta$, we can get an
action of the braid group $B_{n}$ on
$(\Delta)^{n}$. More precisely,
  suppose that
$R : \Delta \times \Delta \rightarrow
\Delta \times \Delta$ is a solution to the
set-theoretic Yang-Baxter equation
from the Lemma above. Let
$\hat{R} =  R \circ \sigma$
with $\sigma : \Delta \times \Delta
\rightarrow
\Delta \times \Delta $
being the  exchange of components,  and let
$ \hat{R}_{i} (n)$ be the endomorphism
of the cartesian power  $\Delta^{n}$
defined by:
\[ \hat{R}_{i} (n) ( (x_{1}, \ldots , x_{n})) =
  (x_{1},   \ldots , x_{i-1},
  \hat{R} (x_{i}, x_{i+1}) , x_{i+2} , \ldots , x_{n} ). \]
Then by the assignment of
$ \hat{R}_{i} (n)$
to the $i^{th}$ generator
$b_{i}$ of the braid group $B_{n}$ we
 obtain an action of $B_{n}$
on $\Delta^{n}$ for each $n$.

In what follows all the examples will be
satisfied the identity
\[ a^{a}=a \ \ \ \ \ \mbox{for all} \ \ \ \
a \in \Delta , \]
which we call the {\bf quandle} condition.
This condition is quarantine the first
Reidemeister move.
We shall call a rack satisfying the
quandle condition a {\bf quandle rack}
or {\bf quandle}.
The term quandle is due to Joyce \cite{Joyce}.

Finally, we recall a definition
of Freyd and Yetter \cite{FY}
(see also \cite{FR}).

\begin{defin}
A (right)  crossed $G-$set for a group $G$
is a set, X, with a right action
of the group $G$, which we write
\[  (x,g) \mapsto x \cdot g
\ \ \ \ \mbox{where} \ \ \ \ x,x \cdot g
\in X \ \ \ \mbox{and} \ \ \ g \in G\]
and a function $\delta : X \rightarrow
G$ satisfying the
{\bf augmentation identity}:
\[  \delta (a \cdot g) =
g^{-1} (\delta (a)) g \ \ \
\mbox{for all} \ \ \ a \in X, g \in G , \]
which is precisely the same as saying that
$\delta$ is $G$-map when $G$ is regarded
as a right $G$-set under right conjugation.
\end{defin}

Given a crossed $G$-set $X$, we can define
an operation of$X$ on itself by defining
$a^{b}$ to be $a \cdot \delta (b)$.
One can easily check that the operation
$(a,b) \mapsto a^{b}$ gives us the rack
structure on $X$, which we call
{\em the augmented rack with
augmentation $\delta$}.
For more details on the theory of augmented
rack see \cite{FR}.

\section{Quasitriangular Poisson Lie Groups}

Let $G$ be a Poisson Lie group. This means that $G$ is a Lie
group equipped with a Poisson structure $\pi$ such that the
multiplication in $G$ viewed as map $G \times G \rightarrow G $
is a Poisson mapping, where $G \times G $ carries the product
Poisson structure. The theory of Poisson Lie groups is a
quasiclassical version of the theory of quantum groups.

  One can easily check that the Poisson structure $\pi$ must
  vanish at the identity $ e \in G$, so that its linearization
  $d_{e} \pi : \G \rightarrow \G \wedge \G$ at $e$ is well
  defined ( here $\G$ is the Lie algebra of $G$ ). It turns out
  that this linear homomorphism is a 1-- cocycle with respect to
  the adjoint action. Moreover, the dual homomorphism $\G^{\ast}
  \wedge \G^{\ast} \rightarrow \G^{\ast}$ satisfies the Jacobi
  identity; i.e. $ \G^{\ast}$ becomes a Lie algebra. Such a pair
  ($\G , \G^{\ast}$) is called a Lie bialgebra \cite{D1}. Each
  Lie bialgebra corresponds to a unique connected, simply
  connected Poisson Lie group. It is easy to show that the pair
  ($\G^{\ast} , \G$) is a Lie bialgebra as soon as ($\G ,
  \G^{\ast}$) is one. The Poisson Lie group ($G^{\ast},
  \pi^{\ast}$) corresponding to ($\G^{\ast} , \G$) will be called
  dual to ($G , \pi$). Thus ( connected, simply connected )
  Poisson Lie groups come in dual pairs.

$\G$ and $ \G^{\ast}$ may be put as Lie subalgebras into the
greater Lie algebra $\widetilde{\G}$ which is called the double Lie
algebra. A vector space  $\widetilde{\G}$ equals $\G \oplus \G^{\ast}$,
 with  Lie bracket
\[ [X + \xi , Y + \eta ] =
[X,Y] + [\xi , \eta] + ad^{\ast}_{X} \eta - ad^{\ast}_{Y} \xi +
ad^{\ast}_{\xi} Y - ad^{\ast}_{\eta} Y \]

Here  $X, Y \in \G$ , $\xi , \eta \in  \G^{\ast}$  and
$ad^{\ast}$ denotes the coadjoint representations of
$\G$  on  $ \G^{\ast}$ and of   $ \G^{\ast}$ on
$\G =( \G^{\ast})^{\ast}$.
 We use [ , ] to denote both the bracket on $\G$ and $
 \G^{\ast}$.

 With respect to $ad$--invariant non-degenerate
canonical bilinear form
\[ (X + \xi , Y + \eta ) = \langle X,
\eta \rangle + \langle Y, \xi \rangle \]
$\G$ and $ \G^{\ast}$ form maximal isotropic subspaces of
$\widetilde{\G}$.

The simply connected group $\widetilde{G}$ corresponding to
$\widetilde{\G}$ is the classical Drinfeld double of the Poisson
Lie group ($G , \pi$).

Conversely, any Lie algebra $\widetilde{\G}$ with a
non--degenerate symmetric $ad$--invariant bilinear form and a
pair of  maximal isotropic subalgebras ( a Manin triple )
gives a pair of dual Lie bialgebras by identifying one of the
subalgebras with the dual of the other by means of this bilinear
form.

Let $r= \sum_{i} a_{i} \otimes b^{i}$ be an element of $\G
\otimes \G$ ; we say that $r$ satisfies the classical
Yang--Baxter equation if
 \[ [ r_{12}, r_{13} ] + [ r_{12}, r_{23}]
+ [ r_{13}, r_{23} ] = 0 .\]
 Here, for instance, $ [ r_{12},
r_{13} ] = \sum_{i,j} [a_{i} , a_{j}] \otimes b^{i} \otimes
b^{j}$. A quasitriangular Lie bialgebra is a pair ($\G , r$),
where $\G$ is Lie bialgebra, $r \in \G \otimes \G$, the
coboundary of $r$ is the cobracket $d_{e} \pi : \G \rightarrow \G
\wedge \G$ and $r$ satisfies the classical Yang--Baxter equation.

Let us associate with $r$ a linear operator
\[  r_{+} :    \G^{\ast}  \rightarrow \G, \ \ \ \ \
\xi \mapsto \langle r, \xi \otimes id \rangle .\]
Its adjoint is given by
\[  -r_{-} = r_{+}^{\ast}:    \G^{\ast}  \rightarrow \G, \ \ \ \ \
\xi \mapsto \langle r, id \otimes \xi \rangle  =
\langle P(r), \xi \otimes id \rangle ,\]
where $P$ is the  permutation operator in $\G \times \G$,
$P(X \otimes Y) = Y \otimes X . $

The Lie bracket  [,] in $ \G^{\ast}$ is given by
\[   [ \xi , \eta ] = ad_{r_{+}(\xi)}^{\ast} \eta  -
ad_{r_{-}(\eta)}^{\ast} \xi\]

\begin{lemma}[\cite{WX}, \cite{RS} ]

For any quasitriangular Lie bialgebra ($\G , r$), the linear maps
\[   r_{+}, r_{-} : \G^{\ast} \rightarrow \G,  \]
defined above, are both Lie algebra homomorphisms.

\end{lemma}
We now turn our attention to groups.

\begin{defin}[\cite{WX}]

A Poisson Lie group $G$ is called quasitriangular if its
corresponding Lie bialgebra $(\G , \G^{\ast})$ is quasitriangular
and if the Lie algebra homomorphisms $ r_{+}$ and $ r_{-}$ from
$\G^{\ast}$ to $ \G$ lift to Lie group homomorphisms $R_{+}$ and
$R_{-}$ from $G^{\ast}$ to $G$. \end{defin}

It turns out that if $G$ is quasitriangular, the maps $\phi$ and $\psi$
from   $G^{\ast}$ to $G$  are Poisson
morphisms, where $\phi (x) = R_{+} (x^{-1}), \  \psi
   (x) = R_{-} (x^{-1}) $,
for any $x \in  G^{\ast}$.
For every Poisson Lie group $G$ there
 are naturally defined left and
right ``dressing'' actions of $G$ on $G^{\ast}$ \cite{STS} ,
whose orbits are
exactly the symplectic leaves of $G^{\ast}$. When  $G$ has the zero
Poisson structure, its dual Poisson Lie group is simply $\G^{\ast}$
with the abelian Lie group structure and ordinary Lie--Poisson bracket.
The left and right dressing actions in this case  are simply the left
and right coadjoint actions of $G$ on  $\G^{\ast}$.

Given a Poisson Lie group $(G, \pi)$, we can consider the Lie algebra
homomorphism from $\G$ to the Lie algebra of vector fields on $G^{\ast}$.
To  describe this homomorphism, we pick an element
$ v \in \G = (\G^{\ast})^{\ast}. $ It may be identified with an element
$\alpha_{v} \in  T_{e}^{\ast} G^{\ast}$.
 Let $\alpha$ be an extension of  $\alpha_{v} $
  to a right-invariant 1-form
on $G^{\ast}$. Then the vector field $v_{\ast}$ corresponding to $v$
is obtained from $- \alpha$ by means of the
Poisson structure $\pi^{\ast} $.
It turns out that in this way   we get a Lie
 algebra homomorphism from
$\G$  to the Lie algebra of vector fields on $\G^{\ast}$ \cite{W}.
Hence, if all the vector fields $v_{\ast}$ are complete ($G$ is
a complete
Poisson Lie group), we obtain a $G$--action $\lambda$  on
$G^{\ast}$  called the
left dressing action. If in this construction
we replace the right-invariant
1-form by the left-invariant 1-form
on $G^{\ast}$ we obtain the right dressing action $\rho$ of $G$
on $G^{\ast}$.

\section{Poisson Lie Rack}
Let $G$  be a quasitriangular
Poisson Lie group
with Lie bialgebra $\G$,
and let $G^{\ast}$ be its
simply connected dual.
We can
lift the Lie algebra homomorphisms
$r_{\pm} : \G^{\ast}  \rightarrow \G$
to the  group homomorphisms
 $R_{\pm} : G^{\ast}  \rightarrow G$,
and define the map
 $ J : G^{\ast}  \rightarrow G$
by $J(x) = R_{+} (x) (R_{-}(x))^{-1} $.
The group $G$ is {\em  factorizable}
if $J$ is a global diffeomorphism.
In this case for each element $x \in G$
we have a factorization
$x = x_{+}  x _{-}^{-1}$,
where  $x_{\pm} = R_{\pm} (J^{-1} (x))$.

The following Theorem
gives us
an augmented rack
structure for $G^{\ast}$.

\begin{th} Let $G$  be a quasitriangular
Poisson Lie group
and let $G^{\ast}$ be its
simply connected dual. Then
$G^{\ast}$ has a structure of crossed $G$-set
with (right) action
\[ (x,g) \mapsto \lambda_{g^{-1}} x  \]
and augmentation $\delta$:
\[ \delta (x) = \phi (x^{-1}) \psi (x)  .\]

\end{th}

This Theorem is the generalization
for  quasitriangular
Poisson Lie group of the
 well-known fact that the left dressing
action of $G$  on $G^{\ast}$, for factorizable Poisson
Lie group $G$, coincides with the conjugation action
$Ad_{x}$ under
the identification of  $G$  with $G^{\ast}$ by $J$.

{\it Proof.}  It follows from  Lemma 2.1
that  the Lie algebra homomorphisms
$r_{\pm}$ naturally extend to
Lie  algebra homomorphisms
$f_{\pm}$ from the double Lie algebra
$\widetilde{\G}$ onto $\G$ defined by:
$f_{\pm} (X + \xi) = X + r_{\pm} \xi$.
By $F_{\pm}$ we denote the Lie group homomorphism
from the classical Drinfeld double
$\widetilde{G}$  of the Poisson
Lie group ($G , \pi$).
For any $d=gu=\bar{u}  \bar{g} \in \widetilde{G}$
with $g, \bar{g} \in G$ and $u, \bar{u} \in G^{\ast}$,
we have \[ F_{+}=g \phi (u^{-1}) =
\phi (\bar{u}^{-1}) \bar{g} , \]
and
\[ F_{-}=g \psi (u^{-1}) =
\psi (\bar{u}^{-1}) \bar{g} . \]
This implies that
\[  \phi (\bar{u}^{-1}) \psi (\bar{u}) =
 g \phi (u^{-1}) \psi (u) g^{-1}.\]
Finally, a straightforward calculation
based on the identity $\lambda_{g} u = \bar{u}$
shows that
$G^{\ast}$ has a structure of crossed $G$-set.
Q.E.D.

It follows from  Weinstein-Xu
result (Lemma 8.5 from \cite{WX})         that
the rack from Theorem 3.1 is the quandle.

\begin{defin}
Let $G$  be a quasitriangular
Poisson Lie group
and let $G^{\ast}$ be its
simply connected dual.
The {\bf Poisson Lie quandle}
is $G^{\ast}$  with the following
rack operation:
\[ b^{a} = \lambda_{\psi(a^{-1}) \phi (a)} b .  \]

Symplectic Poisson Lie quandles are the
symplectic leaves of $G^{\ast}$
( orbits of ``dressing'' actions of $G$
on $G^{\ast}$ ) with the same rack operation.
\end{defin}

For any rack  $\Delta$
we have braid group $B_{n}$
action on $\Delta^{n}$.
Recall  that two braids
give rise to equivalent links if and only
if they are equivalent under Markov moves.
There are two types of Markov moves:
one is conjugation $A  \rightarrow BAB^{-1}$;
the other is by increasing the
number of strings in braid by
a simple twist:
$A \rightarrow A b_{n}^{\pm}$,
for $A \in B_{n}$, where $b_{n}$
is $n^{th}$ generator of $B_{n+1}$.
After an elementary calculation we get

\begin{lemma}
Suppose that  $\Delta$ is a quandle.
 If $A \in B_{n}$
and $B \in B_{m}$ define equivalent
links, then the fixed point sets
of $A$ on $\Delta^{n}$ and of $B$
on $\Delta^{m}$ are isomorphic.

\end{lemma}

This implies that for the
Poisson Lie quandle $G^{\ast}$ we have
link invariant as the fixed point set
of the corresponding braid
action on $(G^{\ast})^{n}$.
These invariant is the space
of the representations of
the fundamental augmented rack of
a link into Poisson Lie quandle
(see Proposition 7.6 from \cite{FR}).

It turns out that this space
of the representations of
the fundamental augmented rack of
a link into Poisson Lie quandle
equal to the Weinstein-Xu link
invariant, associated with
the quasitriangular Poisson Lie group
$G$ \cite{WX}. This will be proofed in the
next paper.

{\bf Problem 1.} Find for the Poisson Lie
quandle the analogue of the
``exchenge'' and of the
compatibility conditions for quantum
R-matrix (see Theorem 5.4 from \cite{WX}).

It is well-known (S. Majid \cite{Majid}) that
a vector-space with a basis having the structure of crossed
$G$-set is equivalent to a representation of the
quantum double of the qroup algebra $C[G]$
with the usual Hopf algebra structure.

{\bf Problem 2.} Find
the quantum analogue of the rack operation on
$G^{\ast}$.


\end{document}